\documentclass[twoside,11pt]{article}

\usepackage{jmlr2e}
\usepackage{comment}
\usepackage{graphicx}
\usepackage{float}
\usepackage{subcaption}
\usepackage{booktabs}
\usepackage[bottom]{footmisc}
\usepackage{hhline}

\usepackage{amsmath}
\usepackage{booktabs}
\usepackage{cmap}
\usepackage{comment}
\usepackage{enumitem}
\usepackage{graphicx}
\usepackage{hhline}
\usepackage{array} 
\usepackage{subcaption}
\usepackage{tcolorbox}
\usepackage[T1]{fontenc}
\usepackage{url}

\ShortHeadings{Racial Disparities and Mistrust in End-of-Life Care}{Boag}
\firstpageno{1}

\begin{document}

\title{Racial Disparities and Mistrust in End-of-Life Care}

\author{\name Willie Boag \email wboag@mit.edu \\
          \addr MIT CSAIL\\
          Cambridge, MA. USA 
          \AND
       \name Harini Suresh \email hsuresh@mit.edu \\
          \addr MIT CSAIL\\
          Cambridge, MA. USA 
          \AND
       \name Leo Anthony Celi, MD, MPH \email lceli@bidmc.harvard.edu\\
          \addr MIT IMES\\
          Cambridge, MA. USA 
          \AND
       \name  Peter Szolovits, PhD  \email psz@mit.edu \\
          \addr MIT CSAIL\\
          Cambridge, MA. USA 
          \AND 
       \name Marzyeh Ghassemi, PhD \email marzyeh@cs.toronto.edu \\
          \addr University of Toronto, Vector Institute\\
          Toronto, ON. CA }

\maketitle

\begin{abstract}
There are established racial disparities in healthcare, including during end-of-life care, when poor communication and trust can lead to suboptimal outcomes for patients and their families.
In this work, we find that racial disparities which have been reported in existing literature are also present in the MIMIC-III database. We hypothesize that one underlying cause of this disparity is due to mistrust between patient and caregivers, and we develop multiple possible trust metric proxies (using coded interpersonal variables and clinical notes) to measure this phenomenon more directly. These metrics show even stronger disparities in end-of-life care than race does, and they also tend to demonstrate statistically significant higher levels of mistrust for black patients than white ones.
Finally, we demonstrate that these metrics improve performance on three clinical tasks: in-hospital mortality, discharge against medical advice (AMA) and modified care status (e.g., DNR, DNI, etc.).
\end{abstract}



\section{Introduction}

There are well-established gaps in the American healthcare system for minority populations: white Americans live 4.5 years longer than African Americans \citep{nchs:life-expectancy-gap}, infant mortality rates are twice as high for African Americans even after adjusting for socioeconomic status \citep{racial-infant-mortality}, African American men are 50\% more likely to develop prostate cancer than white men and are twice as likely to die from it \citep{racial-prostate-cancer}. Differences in care also persist during end-of-life \citep{muni:race-ses-eol-icu,lee:dying-icu}. 

Previous work has suggested that some treatment and outcome disparities are related to higher levels of mistrust: some African Americans have shown suspicion of the clinical motives in advance directives and do-not-resuscitate (DNR) orders \citep{wunsch2010three}, and some report that they suspect that the healthcare system was limiting which treatments they could receive \citep{perkins2002cross}. Mistrust between patient and clinician can be detrimental to patient care, especially in end-of-life (EOL), when patients might defy physician recommendations and insist on higher levels of aggressive care. This would be especially problematic because aggressive EOL care can lead to painful final moments and may not improve patient outcomes \citep{cipolletta2014good}. 
%
%
%
In this work, we turn to a novel source for estimating a patient and clinician's trust relationship: clinical notes and documented interpersonal features. Prior work has established the importance of notes in prediction tasks \citep{ghassemi2014unfolding,ghassemi2015multivariate,suresh2017clinical}, but not in quantifying mistrust. 
Further, while others have worked to establish variations in care in private datasets \citep{pritchard:preferences-death,levinsky:age-expenditures-eol,gessert:race-eol-feeding-tubes}, we target the publicly-available MIMIC dataset~\citep{johnson2016mimiciii} to emphasize reproducibility and encourage future work in this area \footnote{We note that further analysis is available in \cite{boag-eol-mistrust}.}.

We provide three contributions:
\begin{itemize}
\item We quantify the racial disparities in EOL care in the publicly-available MIMIC dataset, and make data extraction and modeling code available for reproduction.\footnote{\url{https://github.com/wboag/eol-mistrust}}
\item We propose multiple proxy trust scores to study the inherent mistrust that patients have in clinical staff providing their care.
\item We demonstrate that our trust scores improve performance in three care-based classification tasks: in-hospital mortality, leaving the hospital against medical advice (AMA), and care status (e.g. Do Not Resuscitate). 
\end{itemize}

\section{Background and Related Work}

\subsection{Medical Treatment Gap}
There are well-established biases in clinical care that follow social biases. For instance, women~\citep{hoffmann2001girl} and obese patients~\citep{phelan2015impact} often have worse treatment options and worse outcomes. Racial disparities have been demonstrated in many care settings; African Americans are less likely to receive pain medication than their white counterparts, even when controlling for covariates such as age, sex, and time of treatment~\cite{jama:pain-meds-kids,plos:pain-meds}. Doctors are more likely to diagnose African Americans with more severe disorders (e.g., schizophrenia and other psychotic disorders), yet less likely to diagnose them with depression \cite{simon:mdd-race,adebimpe:psych-diagnose-black,ruiz:schizophrenia}. 
Biases are especially troubling when applying machine learning because the model might identify and exacerbate biases in a feedback loop \citep{ensign2017runaway}.
However, some select clinical tasks do benefit from knowing the patient's race (e.g. when there are differences in recommended care by genetic makeup). In such settings, race should not be ignored, but algorithms must take steps to reduce unnecessary bias.

\subsection{End-of-Life Care Differences}
During EOL care, minorities are more likely to receive high-intensity, life-sustaining treatments \citep{pritchard:preferences-death,levinsky:age-expenditures-eol,gessert:race-eol-feeding-tubes} and have fewer advance directives\footnote{An advance directive is a legal document that can help ensure patient preferences for various medical treatments are followed if they become incapable of making healthcare decisions.} \citep{smith:acp}. 
White patients are more likely to utilize hospice care and are less likely to disenroll in it than non-white patients \citep{garrett:lifesustaining,hopp:race-eol}. While some of these differences may be attributed to cultural preferences, many issues can also be the result of poor communication or unclear expectations. These imbalances are potentially harmful because aggressive care does not always lead to improved patient outcomes \citep{cipolletta2014good}.

\subsection{Medical Mistrust Among Minority Communities}\label{sec:mistrust-background}
Recent work has explored the multi-faceted history of mistrust between the African Americans community and medical institutions \citep{medicalapartheid}. Poor trust can specifically impact end-of-life care; family members of African American patients are more likely to cite absent or problematic communication with physicians about EOL care \citep{hauser:advance-directives}. Similarly, surveyed African Americans report lower rates of satisfaction with received quality of care~\citep{hanchate:minorities-cost-more}. 
Mistrust can potentially help understand racial disparities in aggressive treatments such as mechanical ventilation and vasopressors. When further invasive procedures are unlikely to succeed or return the patient to a normal lifestyle, doctors may recommend withdrawing treatment and transitioning to comfort-based measures to ensure the patient does not suffer. However, mistrust may lead a patient or healthcare proxy to question the intention of the assessment (e.g. the hospital doesn't want to use resources), and instead demand more aggressive interventions \citep{garrett:lifesustaining,hopp:race-eol}. 

\subsection{Quantifying Trust}

Trust is shaped by subtle interactions in perceived discrimination, racial discordance, poor communication, language barriers, unsatisfied expectations, cultural stigmas and reputations, and is therefore difficult to quantify~\citep{cultural-mistrust}. However, trust is a crucial part of medical care; increased levels of doctor-patient trust have been associated with stronger adherence to a physician's advice and improved health status \citep{trust-improves-outcomes}. Previous efforts to create trust-based measures that correlate with outcomes have relied on surveys, which can be difficult to conduct for both methodological (e.g., selection bias) and practical (e.g., de-identification) reasons \citep{lee:dying-icu}.

\section{Data}
\subsection{Data Source}
\label{sec:data}

\begin{table}
\caption{Population characteristics by race. Parenthetical numbers for categorical variables denote \% membership. Bracketed numbers for continuous variables denote 95\% confidence intervals.}
\hspace{-10mm}
\begin{tabular}{|l|r|rr|r|}
\toprule
\textbf{Variable} & \textbf{Value} &      \textbf{Black}               &       \textbf{White}       & \textbf{p-value}        \\ \hhline{=====}
Population Size & {} &                 1214 &                 9987 & -- -- -- -- -- \\ 
\midrule
Insurance & Private  &      141 (11.61\%) &  1594 (15.96\%)   & $<0.001$  \\
          & Public   &     1062 (87.48\%) &  8356 (83.67\%)   &  \\
          & Self-Pay &  11 ($\;\;$0.91\%) &  37 ($\;\;$0.37\%)& \\
\midrule
Discharge Location & Deceased                 & 401 (33.03\%) & 3869 (38.74\%) & $<0.001$ \\
                   & Hospice                  &  40 ($\;\;$3.29\%) & 421 ($\;\;$4.22\%) & \\
                   & Skilled Nursing Facility & 773 (63.67\%)      & 5697 (57.04\%)     & \\ 
\midrule
Gender & F &  733 (60.38\%) & 5012 (50.19\%) & $<0.001$ \\
       & M &  481 (39.62\%) & 4975 (49.81\%) & \\
\midrule
Length of stay & {} &   13.90 [5.55,19.56] &   14.08 [6.45,19.45] & 0.222 \\
\midrule
Age & {} &  71.31 [60.21,80.36] &  77.87 [66.61,84.93] & $<0.001$ \\
\bottomrule \end{tabular}
\label{tab:tableone-race}
\end{table}

We use the MIMIC-III v1.4~\citep{johnson2016mimiciii} database, consisting of de-identified EHR data from over 58,000 hospital admissions for nearly 38,600 adult patients. The data was collected from Beth Israel Deaconess Medical Center from 2001--2012. 

We define two cohorts from MIMIC: \texttt{EOL} (11,000 admissions) and \texttt{ALL} (50,000 admissions). We use the EOL cohort to replicate racial disparities in MIMIC that other studies have observed \citep{leo:review}. We use the ALL cohort to develop a metric to score the signs of mistrust in each patient's hospital stay and to predict non-end-of-life clinical tasks. 
Both our data extraction and modeling code are made available to promote reproducibility and further study \citep{johnson:mlhc17-reproduce}.

A patient is added to the EOL cohort if they have a hospital stay which lasted at least 6 hours and they either died in the hospital, were discharged to hospice, or were discharged to a skilled nursing facility\footnote{See the extended masters thesis \citep{boag-eol-mistrust} for additional analysis. Disparities are also replicated on the eICU database, and all results are also performed on a stricter-but-smaller cohort that excludes skilled nursing facilities.}.
Table~\ref{tab:tableone-race} displays summary statistics of the cohort by race. 
A $\chi^2$ test shows significant differences for insurance type, discharge location, and gender ($p<0.001$ for all three). In particular, we see that the black population has both higher rates of uninsurance and publicly-funded insurance than their white counterparts. In lieu of other coded data, this often serves as a proxy for socio-economic status. In addition, white patients have higher in-hospital mortality and hospice rates, whereas a larger percent of black patients are discharged to skilled nursing facilities. Finally, there is a large difference between the black gender ratio (60-40 women) and white gender ratio (50-50).
Using the Mann-Whitney test, the two populations have comparable lengths-of-stay (p=0.222), but significantly different population ages ($p<0.001$).

\subsection{Treatment Extraction}

The main focus for this work is measuring disparities in aggressive end-of-life procedures, so we extract treatment durations (in minutes) from MIMIC's derived mechanical ventilation (\texttt{ventdurations}) and vasopressor (\texttt{vasopressordurations}) tables. Due to the noisiness of clinical measurements -- for instance, when one treatment span is erroneously coded as two back-to-back smaller spans -- we merge any treatment spans that occurred within 10 hours of each other.\footnote{This heuristic was suggested by MIMIC staff because 10 hours is approximately the shift of a nurse, and treatment duration events might get recorded once at the beginning of each shift.} If a patient had multiple spans, such as an intubation-extubation-reintubation, then we consider the patient's treatment duration to be the sum of the individual spans.

\subsection{Patient-Provider Interaction Extraction}

In this work, we quantify the patient's interactions with their nurses and doctors using two sources: clinical notes and coded chart events. 
We obtain the notes of any patient who had a stay at least 12 hours in the ICU. This resulted in 48,273 admissions and over 800,000 notes.
Throughout a patient's stay, caregivers write narrative prose notes to document administered care, record patient preferences, issue reminders and warnings, and comment on the patient's quality of care.
In documenting their impressions, caregivers can give clues into the level of trust in their relationship with their patient. 
To supplement this narrative prose, we also extract coded information from the MIMIC \texttt{chartevents} table. Table \ref{tab:chartevents-features} shows the \texttt{chartevents} information types, with categories including: indication of family meetings, patient education, whether the patient needed to be restrained, how thoroughly pain is being monitored and treated, healthcare literacy (e.g. whether the patient has a healthcare proxy), whether the patient has a support system (such as family, social workers, and religion), and Riker-SAS \citep{riker} and Richmond-RAS \citep{richmond} agitation scales.

\begin{table}
  \begin{center}
    \caption{Coded interpersonal feature types from chartevents.}
    \label{tab:chartevents-features}
    \hspace*{-12mm}
    \begin{tabular}{|c|c|c|c|}
    	\hline
 1:1 sitter present? & baseline pain level (0 to 10) & received bath?  & bedside observer  \\ \hline
 behavioral intervent  & currently experiencing pain  & disease state   & consults  \\ \hline
 education barrier  & education learner  &  education method  & feamily meeting?  \\ \hline
 education readiness  & harm by partner?  &  education topic  & judgement  \\ \hline
 follows commands?  & family communication method   &  gcs - verbal response  & informed?  \\ \hline
 hair washed?  & goal richmond-ras scale  & headache?   & health care proxy?  \\ \hline
 pain management  & non-violent restraints?  & orientation   &  pain (0 to 10) \\ \hline
 pain assess method  & understand \& agree with plan?  & pain level acceptable?   & reason for restraint  \\ \hline
 restraint device  & richmond-ras scale (-5 to +4)   & rsbi deferred  & riker-sas scale  \\ \hline
 safety measures  & violent restraints ordered?  &  security  & security guard  \\ \hline
 side rails  & status and comfort  &  sitter  & skin care?  \\ \hline
 spiritual support  & behavior during application  & support systems   & stress  \\ \hline
 verbal response  & teaching directed toward & wrist restraints?  &  social work consult?   \\ \hline
    \end{tabular}
  \end{center}
\end{table}

\section{Methods}

\subsection{Quantifying Racial Disparities in End-of-Life Care}

We aim to replicate previous findings of racial disparities using MIMIC-III. We take as reference a set of three recent papers which examined the racial disparities in end-of-life care for non-white or minority populations \citep{ices:immigration,muni:race-ses-eol-icu,lee:dying-icu}. We compared the median differences between white and black populations using Mann-Whitney analysis. In accordance with prior work, we consider p-values of $<.05$ to be statistically significant. 


\subsection{Establishing a Medical Mistrust Metric}\label{mistrust-methods}
We aim to quantify trust in the doctor-patient relationship. 
Because it is novel to study the effects of algorithmically-derived trust scores, we employ three metrics to avoid the impression that any single one tells the whole story. Much like fairness, trust may eventually prove to be impossible to fully characterize with a single score \citep{formalizing-fairness}. Either way, we recommend that these scores be taken as a collective proxy for further-refined notions of trust. Two of the scores come from training a model to predict trust-associated labels using interpersonal doctor-patient features, and the third score is out-of-the-box sentiment analysis of the patient's clinical notes.

\begin{figure}
  \caption{An example of a nursing note documenting mistrust (in red). Situation-specific identifying information has been blacked out.}
  \hspace*{22mm}
  \includegraphics[width=105mm]{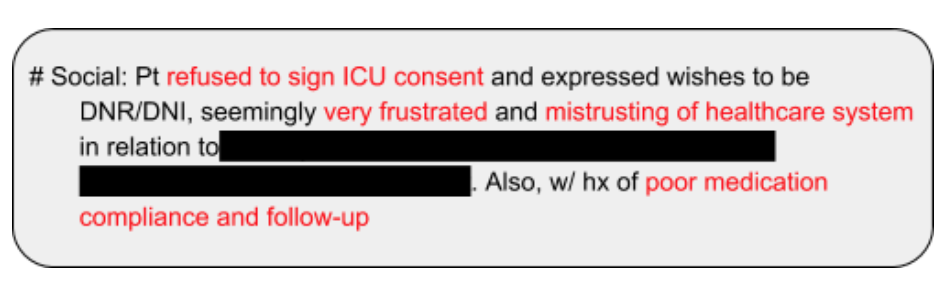}
  \label{img:mistrust-note}
\end{figure}

For the first two metrics, we use an L1-regularized logistic regression to predict labels derived from clinical notes using structured event data. The features in this predictive framework were extracted for the ALL cohort from the \texttt{chartevents} table, as shown in Table \ref{tab:chartevents-features}. In total, we extract 620 binary features.

Figure \ref{img:mistrust-note} shows an example of the various signals of mistrust that are documented in nursing notes. We define two sets of trust-associated labels to train the above classifier using simple rule-based searches to detect:
\begin{enumerate}
    \item \textbf{Noncompliance}: Noncompliance (e.g. refusing to adhere to follow-ups, regiments, take medicine, etc) indicates a very overt mistrust; rather than just holding an unspoken resentment, the patient actually defies their doctor's orders. It explicitly demonstrates that the patient is willing to disregard physician decisions
    \item \textbf{Autopsy}: One obvious benefit of an autopsy is quality assurance: did patient receive the proper treatment for the proper disease? Often times, families decline autopsies because they feel that dissecting a loved one would not be worth it when they trust that the doctor did everything they could. Conversely, higher autopsy rates could indicate patients suspect the doctor made a mistake. In this dataset, black patients (38.5\% autopsy rate) are autopsied much more often than white patients (24.3\% autopsy rate).
\end{enumerate}
Once these model are trained, we use each classifier's predicted probability as a measure of their mistrust for a new patient. This score defines a spectrum of trust based on how many indicators the patient has that are associated with typically poor interactions vs typically good interactions.

To encourage diversity of possible metrics, the third notion of mistrust is defined as the negative of the sentiment score from the patient's clinical notes in the caregiver's own words.
We use the Pattern software package \citep{pattern} and concatenate all of the notes from the stay into one document and tokenized using whitespace as a delimiter.\footnote{This step is actually important because a naive application of tokenization results in even positive notes which contain identified tags like ``Date:[**5-1-18**]'' to be tagged as negative because the tool's string-matching algorithm was identifying ``:['' as negative emoticon.}

Finally, we normalize each of the mistrust score distributions to be zero-mean and unit-variance, which helps for comparison.

\subsection{Prediction of Downstream Clinical Outcomes}

Trust is vital to a healthy doctor-patient relationship. A mistrustful patient might be reluctant to share sensitive, but potentially important information with their doctor.
To further explore the impact of modeled trust, we examine two trust-associated outcomes (\texttt{Code Status}\footnote{either \texttt{``Full Code''} or \texttt{``DNR / DNI / Comfort Measures Only''}} and \texttt{Whether the patient leave Against Medical Advice (AMA)}) and one more standard outcome (\texttt{in-hospital mortality}). We are interested to see how much value race and trust add as features to a baseline model which uses the patient's age, gender, length-of-stay, and insurance type. We take the average AUCs of 100 runs from randomly chosen 60/40 train/test splits with an L1-regularized logistic regression model.

\section{Results}

\subsection{Racial Treatment Disparities in EOL Care are Significant}
\label{sec:treatment-stats}

\begin{figure}[!h]
\centering
    \caption{We observe racial disparities in aggressive interventions for black patients compared to white patients. Medians are indicated by dotted lines; differences are significant ($p < 0.05$) for ventilation but not for vasopressors.}
    \begin{subfigure}{.40\linewidth}
      \centering
      \includegraphics[width=\textwidth]{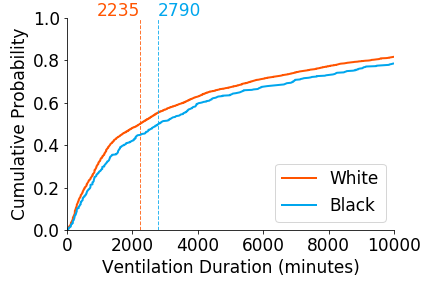}
      \caption{CDF of ventilation \\ duration by race (\textcolor{red}{$p=.005$}). 
      }
    \end{subfigure}
	~
   \vspace{8mm}
    \begin{subfigure}{.40\linewidth}
 		\centering
	 	\includegraphics[width=\linewidth]{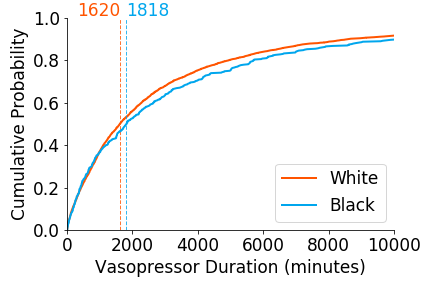}
 		\caption{CDF of vasopressor \\ duration by race ($p = 0.12$).
      }
    \end{subfigure}
    \label{fig:treatments-race}
\end{figure}

We demonstrate racial treatment disparities in the MIMIC dataset do exist. Figure \ref{fig:treatments-race} highlights the differences in white and black populations for aggressive treatment durations. Figures \ref{fig:treatments-race}a and \ref{fig:treatments-race}b show that for both mechanical ventilation and vasopressors, the median black patient receives a longer duration of treatment, suggesting a reluctance to transition to palliative care. While these results only show statistical significance for ventilation (p=0.005), the same trends are also observable for vasopressor administration (p=0.12) . 

\subsection{Quantifying Mistrust Better Captures EOL Treatment Disparities}\label{trust-metric}

\begin{table}[!b]
  \begin{center}
      \caption{Top-3 most positively and negatively informative chartevent features for each mistrust metric.}
    \begin{minipage}{.5\linewidth}
	  \label{tab:metric-weights}
    \begin{tabular}{|c|c|}
    	\hline
                     \textbf{Feature} & \textbf{Weight} \\ \hline \hline
                         state: alert & -1.0156  \\ \hline
                    riker-sas scale:  &  0.7013  \\
                             agitated &          \\ \hline
                           pain: none & -0.5427  \\ \hline
                  richmond-ras scale: & -0.3598  \\  
                     0 alert and calm &          \\ \hline
              education readiness: no &  0.2540  \\ \hline
         pain level: 7-mod to severe  &  0.2168  \\ \hline
    \end{tabular}
    \subcaption{Noncompliance-derived Mistrust}
    \end{minipage}
~
	\begin{minipage}{.33\linewidth}
    \begin{tabular}{|c|c|}
    	\hline
                     \textbf{Feature} & \textbf{Weight} \\ \hline \hline
                       pain present: no & -0.2689  \\ \hline
                        spokesperson is & -0.2271  \\
                       healthcare proxy &          \\ \hline
                  family communication: & -0.1184  \\
                         talked to m.d. &          \\ \hline
                  reapplied restraints  &  0.1153  \\ \hline
              restraint type: soft limb &  0.0980  \\ \hline
               orientation: oriented 3x &  0.0363  \\ \hline
    \end{tabular}
    \subcaption{Autopsy-derived Mistrust}
    \end{minipage}
    ~
  \end{center}
\end{table}

\subsubsection{Creation of a Mistrust Metric}

Table~\ref{tab:metric-weights} shows the three most positively and most negatively informative weights in each model learned while fitting a mistrust metric. 
The features align well with an intuitive notion of mistrust: patients who are agitated, restrained, and unreceptive to education are more likely to be mistrustful. Conversely, we see that calm, pain-free patients with good communication are more willing to trust their doctor.

\subsubsection{Black Patients Have Higher Levels of Mistrust}

We observe statistically significant racial disparities in the two-out-of-three mistrust metrics. For both noncompliance-derived mistrust and negative sentiment, the median black patient has a higher level of mistrust than the median white patient using the Mann-Whitney test
(p=0.003 and p=0.007, respectively). This is not surprising, given the extensive literature investigating differences in iatrophobia by race \citep{medicalapartheid}. Interestingly, there were virtually no racial disparities in the autopsy-derived metric (p=0.13), which is especially unexpected given the higher rate of autopsies among African Americans.

\subsubsection{Trust-based Disparities in End-of-Life Care}

We hypothesize that if trust were a contributing factor to EOL treatment disparities, then stratifying the data into low mistrust and high mistrust\footnote{For each treatment, we preserve the same size difference of stratified groups in order to maintain consistency in sample sizes for significance testing (e.g. since the black group contains 510 patients for ventilation, we compare the 510 lowest trust patients against the 4811 highest trust patients).} would yield an even larger disparity than white and black.

\begin{figure}
\centering
    \caption{\textbf{Noncompliance Cohort Disparities:} A cohort of noncompliance-derived mistrust admissions yields significant differences in both ventilation and vasopressor duration.}
    \hspace{-10mm}
    \begin{subfigure}{.30\linewidth}
      \centering
      \includegraphics[width=\textwidth]{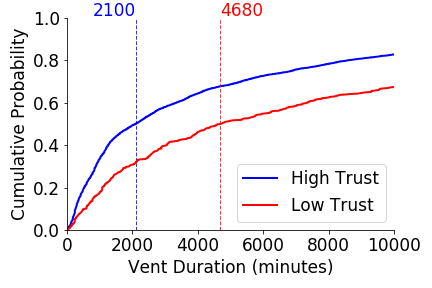}
      \centering
      \caption{\textbf{Mechanical Ventilation}\\
          \textbf{High Trust}: 4810 patients\\
          \textbf{Low Trust}: \;\;\;510 patients\\
          \textcolor{red}{ $p<0.001$ }}
    \end{subfigure}
	~
    \hspace{10mm}
    \begin{subfigure}{.30\linewidth}
      \centering
      \includegraphics[width=\textwidth]{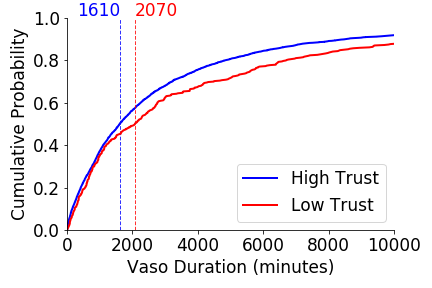}
      \centering
      \caption{\textbf{Vasopressors}\\
          \textbf{High Trust}: 4456 patients\\
          \textbf{Low Trust}: \;\;\;453 patients\\
          \textcolor{red}{p=0.001}}
    \end{subfigure}
    \label{fig:mistrust-noncompliance-treatments}
\end{figure}

Figure \ref{fig:mistrust-noncompliance-treatments} shows significant disparities for both ventilation ($p<0.001$) and vasopressor (p=0.001) durations using the noncompliance-derived metric to stratify the cohort.
The difference between the medians of the mistrustful and trustful groups is 650 minutes for vasopressors (as opposed to 200 minutes between black and white). This gap is even larger for ventilation durations, as shown in Figure \ref{fig:mistrust-noncompliance-treatments}a: the noncompliance stratification shows a 2580-minute difference between medians, in contrast to the 832-minute gap for the race split in Figure \ref{fig:treatments-race}a. This threefold-increase in the treatment gap suggests that trust might be one of the contributing factors for the original racial disparity.

\begin{figure}
\centering
    \caption{\textbf{Autopsy Cohort Disparities:} A cohort of autopsy-derived mistrust admissions yields significant differences in ventilation, but a non-significant difference in vasopressor duration.}
     \hspace{-10mm}
     \begin{subfigure}{.30\linewidth}
       \centering
       \includegraphics[width=\textwidth]{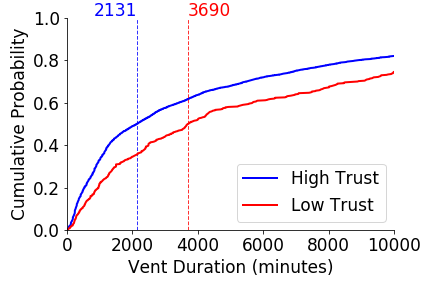}
       \centering
       \caption{\textbf{Mechanical Ventilation}\\
           \textbf{High Trust}: 4810 patients\\
           \textbf{Low Trust}: \;\;\;510 patients\\
           \textcolor{red}{p<0.001}}
     \end{subfigure}
 	~
     \hspace{10mm}
     \begin{subfigure}{.30\linewidth}
       \centering
       \includegraphics[width=\textwidth]{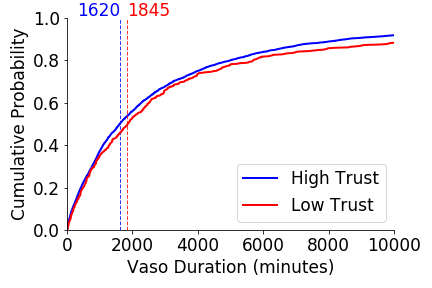}
       \centering
       \caption{\textbf{Vasopressors}\\
           \textbf{High Trust}: 4456 patients\\
           \textbf{Low Trust}: \;\;\;453 patients\\
           p=0.059}
     \end{subfigure}
     \label{fig:mistrust-autopsy-treatments}
\end{figure}

For autopsy-derived mistrust, we can see that Figure \ref{fig:mistrust-autopsy-treatments} exhibits the same conclusions as race-based stratifications in Figure \ref{fig:treatments-race}: mechanical ventilation has significant disparities ($p<0.001$) whereas vasopressors do not (p$=0.059$). However, just as noncompliance-based mistrust had a threefold increase in the treatment gap, this autopsy-derived metric has a twofold increase from the racial disparities found in ventilation (1,559 vs. 832 minutes) and vasopressors (245 vs 106 minutes).

\begin{figure}
\centering
    \caption{\textbf{Sentiment Cohort Disparities:} A cohort of negative sentiment analysis admissions yields significant differences in ventilation, but virtually no differences in vasopressor duration.}
    \hspace{-10mm}
    \begin{subfigure}{.30\linewidth}
      \centering
      \includegraphics[width=\textwidth]{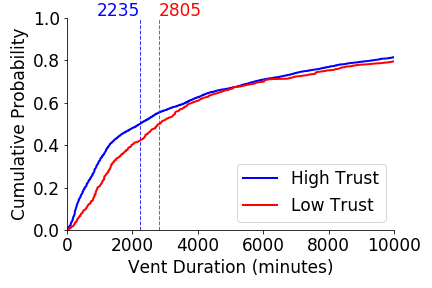}
      \centering
      \caption{\textbf{Mechanical Ventilation}\\
          \textbf{High Trust}: 4646 patients\\
          \textbf{Low Trust}: \;\;\;492 patients\\
          \textcolor{red}{$p<0.001$}}
    \end{subfigure}
	~
    \hspace{10mm}
    \begin{subfigure}{.30\linewidth}
      \centering
      \includegraphics[width=\textwidth]{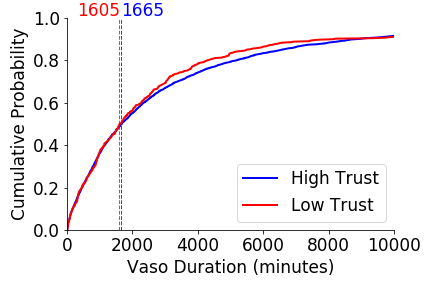}
      \centering
      \caption{\textbf{Vasopressors}\\
          \textbf{High Trust}: 4284 patients\\
          \textbf{Low Trust}: \;\;\;427 patients\\
          p=0.241}
    \end{subfigure}
    \label{fig:neg-sentiment-treatments}
\end{figure}

Negative sentiment analysis exhibits the same trend for ventilation ($p<0.001$) but a surprising result for vasopressor usage, as shown in Figure \ref{fig:neg-sentiment-treatments}. 
There seems to be virtually no sentiment-based difference at all in vasopressor duration (p$=0.241$). In fact, even the ventilation gap is smaller than with the other mistrust-based cohorts: 570 minutes (gaps for noncompliance and autopsy were 2,580 and 1,559, respectively).
These results show that sentiment analysis is a bit of an outlier from the other two mistrust metrics. 
Nonetheless, we believe even this metric's results are a useful contribution for exploring the space of algorithmically-defined trust.

\subsubsection{Not Just Some Acuity Score Proxies}

\begin{table}[b]
  \begin{center}
    \caption{Pairwise Pearson correlation coefficients between scores.\\}
    \label{tab:correlations}
    \hspace*{-1cm}
    \begin{tabular}{|c||c|c|c|c|c|}
    	\hline
              & OASIS & SAPS II & Noncompliance & Autopsy & Sentiment \\ \hline 
\hline
      OASIS   &  1.0   &  0.679 & 0.050  & -0.012 &  0.075 \\ \hline
     SAPS II  &  0.679 &  1.0   & 0.013  & -0.013 &  0.086 \\ \hline
Noncompliance &  0.050 &  0.013 & 1.0    &  0.262 &  0.058 \\ \hline
      Autopsy & -0.012 & -0.013 & 0.262  &  1.0   &  0.044 \\ \hline
    Sentiment &  0.075 &  0.086 & 0.058  &  0.044 &  1.0   \\ \hline
    \end{tabular}
  \end{center}
\end{table}

These mistrust metrics are typically more effective than race at stratifying the data to show treatment disparities.
However, one possible concern is the possibility that the scores are capturing severity-of-illness rather than mistrust: certainly, high-risk patients would be treated differently than the general population. To address this concern, we compare the pairwise Pearson correlation coefficients between the three mistrust scores and two well-established acuity scores (OASIS and SAPS II). Table \ref{tab:correlations} shows that the severity scores have a strong correlation with one another (0.68). On the other hand, every mistrust scores has a very weak of a correlation with these risk scores; the largest severity-mistrust correlation being 0.086 between Sentiment and SAPS II. Interestingly, the autopsy-based mistrust metric is actually negatively correlated with the two severity scores (i.e. sicker patients are less likely to get autopsies) while still remaining positively correlated with the other two mistrust metrics. The noncompliant and autopsy metrics have the strongest intra-mistrust correlation (0.262). This is not surprising because these two metrics are both derived from Machine Learning on the \texttt{chartevents} features.

\subsection{Mistrust Metric is Predictive in Downstream Tasks}

We demonstrate that the mistrust score captures meaningful information by evaluating its contribution as a feature to predict three tasks: whether the patient is DNR, whether the patient left AMA, and in-hospital mortality.

The results can be found in Table~\ref{tab:mistrust-classification-results}, which show that race and trust both improve outcome prediction. Performance is variable across the tasks: no single feature is most useful for all three tasks. As is often the case, combining all of the features achieves the best results on each task --- sometimes even with statistical significance, as for in-hospital mortality. Each mistrust metric achieves the top individual performance on one of the tasks: noncompliance-score for \texttt{Left AMA}, autopsy-score for \texttt{Code Status}, and negative-sentiment-score for \texttt{In-Hospital Mortality}. Race proves itself to be a very useful feature for all three tasks, outperforming at least one of the mistrust metrics in each category. Noncompliance-derived mistrust proves to be the most performant mistrust metric, achieving top-2 results for each task (excluding the ALL run).

\begin{table}
  \begin{center}
    \caption{Effect of race and mistrust features on various binary classification tasks. Performance is measured by AUC and averaged over 100 runs.}
    \label{tab:mistrust-classification-results}
    \hspace*{-1cm}
    \begin{tabular}{|l||c|c|c|}
    	\hline
               \textbf{Features} & \textbf{Left}             & \textbf{Code}             & \textbf{In-Hospital}  \\
                                 & \textbf{AMA}              & \textbf{Status}           & \textbf{Mortality}    \\ 
                                 & (n=48,071)                & (n=39,815)                & (n=48,071)            \\ \hline
       Baseline                  & $        0.859  \pm .014$ & $        0.763  \pm .013$ & $        0.600  \pm .011$  \\ \hline
       Baseline + Race           & $        0.861  \pm .014$ & $        0.766  \pm .014$ & $        0.614  \pm .011$  \\ \hline
       Baseline + Noncompliant   & $\textbf{0.869} \pm .012$ & $        0.767  \pm .013$ & $        0.614  \pm .010$  \\ \hline
       Baseline + Autopsy        & $        0.861  \pm .012$ & $\textbf{0.773} \pm .011$ & $        0.603  \pm .012$  \\ \hline
   Baseline + Negative-Sentiment & $        0.859  \pm .013$ & $        0.765  \pm .014$ & $\textbf{0.615}  \pm .010$  \\ \hline \hline
       Baseline + ALL            & $\textbf{0.873} \pm .012$ & $\textbf{0.782} \pm .012$ & $\textbf{0.635}  \pm .010$  \\ \hline
    \end{tabular}
  \end{center}
\end{table}

Average classifier weights for the Baseline + ALL model are shown in Table \ref{tab:avg-downstream-weights}. The two features most strongly associated with in-hospital mortality were the patient's mistrust scores followed by the patient's age. This is not surprising, because the highest-noncompliance-mistrust quartile has a 13.7\% mortality rate, which is over three times as high as the lowest-noncompliance-mistrust quartile's 4.4\% mortality rate. 

We also observe that noncompliance-derived mistrust, autopsy-derived mistrust, and race:black are the only three features positively associated with leaving the hospital AMA. Noncompliance (average coefficient of .52) is significantly more informative than autopsy and race:black (.01 and .03, respectively).
As we saw in earlier experiments, mistrust is an even stronger indicator than race. In general, however, race tends to be a poor predictor for some of these outcomes because it is too coarse-grained to capture all of the different ways healthcare delivery can fail. For most tasks, the weights of racial features add little predictive value and are zero'd out during training. On the other hand, age is a very powerful predictor of these various outcomes, though not consistently indicating breakdowns in the doctor-patient relationship. While older patients are more likely to expire in-hospital, they are less likely to leave the hospital against medical advice. The mistrust score is the only feature positively associated with each outcome and consistently demonstrates predictive value.

\begin{table}[h!]
      \begin{center}
        \caption{Average regularized weights for BASELINE+ALL model on various tasks.}
        \label{tab:avg-downstream-weights}
        \hspace*{-1cm}
        \begin{tabular}{|l|c|c|c|}
            \hline
          \textbf{feature} & \textbf{Left AMA}       & \textbf{Code Status}   & \textbf{Mortality}    \\ \hline \hline
              noncompliant &  $\;\;\;0.52 \pm 0.09$  & $\;\;\;0.27 \pm 0.04$  & $\;\;\;0.16 \pm 0.03$ \\ \hline
                   autopsy &  $\;\;\;0.01 \pm 0.03$  & $\;\;-0.44  \pm 0.05$  & $\;\;\;0.02 \pm 0.02$ \\ \hline
        negative sentiment &  $\;\;\;0.00 \pm 0.02$  & $\;\;\;0.09 \pm 0.03$  & $\;\;\;0.16 \pm 0.03$ \\ \hline
               race: asian &  $\;\;\;0.00 \pm 0.00$  & $\;\;\;0.00 \pm 0.00$  & $-0.05      \pm 0.03$ \\ \hline
               race: black &  $\;\;\;0.03 \pm 0.12$  & $-0.22      \pm 0.19$  & $-0.53      \pm 0.31$ \\ \hline
            race: hispanic &  $\;\;\;0.00 \pm 0.00$  & $-0.17      \pm 0.21$  & $-0.58      \pm 0.34$ \\ \hline
               race: other &  $-0.15      \pm 0.19$  & $-0.12      \pm 0.17$  & $\;\;\;0.15 \pm 0.30$ \\ \hline
               race: white &  $-0.02      \pm 0.06$  & $\;\;\;0.06 \pm 0.15$  & $-0.26      \pm 0.30$ \\ \hline
     race: native american &  $\;\;\;0.00 \pm 0.00$  & $\;\;\;0.00 \pm 0.00$  & $\;\;\;0.00 \pm 0.00$ \\ \hline
              gender: male &  $\;\;\;0.00 \pm 0.00$  & $-0.85      \pm 1.40$  & $-0.67      \pm 0.99$ \\ \hline
            gender: female &  $-0.40      \pm 0.20$  & $-0.49      \pm 1.39$  & $-0.59      \pm 0.99$ \\ \hline
        insurance: private &  $-1.01      \pm 0.21$  & $-0.94      \pm 0.29$  & $-0.96      \pm 0.95$ \\ \hline
         insurance: public &  $\;\;\;0.00 \pm 0.00$  & $-0.02      \pm 0.28$  & $-0.50      \pm 0.95$ \\ \hline
       insurance: self-pay &  $\;\;\;0.00 \pm 0.00$  & $-0.02      \pm 0.24$  & $-0.21      \pm 0.68$ \\ \hline
			length-of-stay &  $-1.44      \pm 0.37$  & $-0.70      \pm 0.10$  & $\;\;\;0.08 \pm 0.03$ \\ \hline
                       age &  $-2.10      \pm 0.21$  & $\;\;\;0.42 \pm 0.02$  & $\;\;\;0.20 \pm 0.02$ \\ \hline
        \end{tabular}
      \end{center}
\end{table}

\section{Conclusion}

One aim of this work is to establish existing findings of racial disparities in end-of-life care on a publicly available dataset, namely MIMIC III. We demonstrate that black patients receive -- sometimes significantly -- longer durations of invasive treatments and are significantly more likely to leave the hospital against medical advice. Though these trends have been studied in private datasets, we present our analysis on a public dataset and make our code available.

To further investigate this phenomenon, we propose multiple proxy mistrust scores using coded interpersonal data and clinical notes. We find that stratifying patients by trust score, instead of by race, more fully separates those patients who persist in aggressive interventions from those who do not.
We show that the mistrust scores add value to multiple predictive tasks. While the scores partially capture racial differences (the median black patient had higher level of mistrusts than the median white patient), it is multifaceted, and feature analysis suggests that it captures information that agrees with our intuition of relationships and trust.

Medical machine learning is moving forward at an exciting pace; this work is a first step towards creating models that serve everyone, and do not propagate existing disparities in care. 
This work does have shortcomings, most notably that the small datasize likely reduced statistical power and that the mistrust metrics were approximations of an intangible concept. Each metric had its flaws: there are non-trust-based reasons why a patient might be noncompliant (e.g. couldn't afford to refill their prescription), not all autopsies suggest skepticism of the doctor's performance, and although sentiment is a related concept it is ultimately not the same thing as trust. 
We encourage forming inter-disciplinary collaborations among machine learning, healthcare, and social science communities to refine algorithmic notions of mistrust, quantify potential biases and disparities in models, and identify potential factors that lead to patient dissatisfaction.

\section*{Acknowledgments}
The authors would like to thank Matthew McDermott for his input and suggestions. This research was funded in part by 
the National Science Foundation Graduate Research Fellowship Program under Grant No. 1122374 as well as the National Institute of Health under grants NIBIB Grant R01 EB017205, U54-HG007963, R01-EB017205, and 1R01MH106577 (National Institute of Mental Health).

\bibliography{mlhc}

\end{document}